\documentclass[pra, aps, twocolumn, superscriptaddress, nofootinbib, showpacs]{revtex4-1}

\usepackage{natbib}
\usepackage{slashed}
\usepackage{graphicx}
\usepackage[usenames, dvipsnames]{color}
\usepackage{hyperref}
\usepackage{bm}
\usepackage{amsmath}
\usepackage{color}

\hypersetup{backref,
colorlinks=true,
linkcolor=blue,
linktoc=page,
citecolor=blue,
urlcolor=blue}{}{}
\usepackage{cleveref} 
\usepackage{cancel}

\everymath{\displaystyle}

\begin{document}

\title{Towards a kinetic theory of a dark soliton gas in one-dimensional superfluids}

\author{Clara Pereira}
\email{clara.pereira@ist.utl.pt}
\affiliation{Instituto de Plasmas e Fus\~ao Nuclear, Lisboa, Portugal}
\affiliation{Instituto Superior T\'ecnico, Lisboa, Portugal}

\author{Jo\~ao D. Rodrigues}
\affiliation{Physics Department, Blackett Laboratory, Imperial College London, United Kingdom}

\author{José T. Mendon\c{c}a}
\affiliation{Instituto de Plasmas e Fus\~ao Nuclear, Lisboa, Portugal}
\affiliation{Instituto Superior T\'ecnico, Lisboa, Portugal}

\author{Hugo Ter\c cas}
\email{hugo.tercas@tecnico.ulisboa.pt}
\affiliation{Instituto de Plasmas e Fus\~ao Nuclear, Lisboa, Portugal}
\affiliation{Instituto Superior T\'ecnico, Lisboa, Portugal}

\begin{abstract}

Soliton hydrodynamics is an appealing tool to describe strong turbulence in low-dimensional systems. Strong turbulence in quasi-one dimensional spuerfluids, such as Bose-Einstein condensates, involves the dynamics of dark solitons and, therefore, the description of a statistical ensemble of dark-solitons, i.e. soliton gases, is necessary. In this work, we propose a phase-space (kinetic) description of dark-soliton gases, introducing a kinetic equation that is formally similar to the Vlasov equation in plasma physics. We show that the proposed kinetic theory can capture the dynamical features of soliton gases and show that it sustains an acoustic mode, a fact that we corroborate with the help of direct numerical simulations. Our findings motivate the investigation of the microscopic structure of out-of-equilibrium and turbulent regimes in low-dimensional superfluids.

\end{abstract}
\maketitle

%%%%%%%%%%%%%
\section{Introduction}

Unlike its classical analogue, which finds a comprehensive model in the Navier-Stokes equation, quantum turbulence (QT) does not fin in a unified and comprehensive description \cite{2014Barenghi}. The difficulties of establishing a suitable framework are rooted not only on the coexistence of normal and superfluid phases, but most relevantly on the topological nature of turbulent structures: while in classical systems they assume arbitrary shapes and sizes,  with lengths spanning over several orders of magnitude, in the quantum regime vorticity is quantised \cite{2019Alperin,2014Barenghi}. As such, QT exhibits vortex tangling, as envisioned by Feynman, resulting in a vorticity distribution that is quite distinct from the continuous vorticity present in classical fluids, making the dynamics of QT rather complicated \cite{Skrbek_2011}. Both theoretical and experimental methods have been developed to produce and investigate vortex tangling in Bose-Einstein condensates (BEC) \cite{yukalov_2016, castilho_2019}. The investigation of the statistical properties of such systems has revealed that vortex tangling can, under certain circumstances, be at the origin of ``classicality", such as the Kolmogorov scaling of the energy spectrum \cite{barenghi_PNAS_2014}.

The features of quantum turbulence change significantly due to dimensional constraints. In two-dimensional (2D) systems, QT shares with classical turbulence an intriguing feature: the so-called inverse cascade, where energy flows backwards from smaller to larger scales, in opposition to what is dictated by Kolmogorov cascade. A consequence of the inverse cascade is the vortex clustering observed in recent experiments \cite{gauthier_2019,johnstone_2019, madeira_2020,griffin_2020, andrew_2020}, as first predicted in Onsager’s theory \cite{Onsager_1949, simula_2014}. The situation is even more drastic in one-dimensional (1D) systems, since quantum fluctuations may play quite a significant role \cite{giamarchi_book, Dziarmaga_2006, edler_2017}. Moreover, angular momentum quantization is not possible in one-dimensional systems, and the role of vortices is played by dark solitons (DS), the topological defects created by a phase jump in the order parameter \cite{frantzeskakis_2010, ku_2016}. Another interesting aspect of dark-solitons is the fact of being fermionic \cite{sato_2012, karpiuk_2015}, being intimately related to the type-II excitations on top of a one-dimensional Bose gas as predicted by the Lieb-Liniger theory \cite{lieb_1963}. Moreover, the concept of ``solitonic turbulence" is also present in some conditions of strong turbulence in classical systems \cite{hassaini_2017, cazaubiel_2018}, which increases the interest around the development of statistical methods for solitons. Previous studies in 1D QT indeed exist, but are mostly (if not exclusively) performed in the weak turbulence regime \cite{nazarenko_2001}. Similarly to what happens in 2D, Zakharov’s weak (or wave--wave) turbulence theory reveals the occurrence of inverse cascades in 1D \cite{connaughton_2005, sun_2012}, but it is still not clear what is supposed to happen in a strong turbulence situation. Can we expect to observe the same sort of inverse cascades for dark solitons in 1D? If yes, how will their fermionic statistics work \cite{tercas_2013, tercas_2014}? \par

With the aim of understanding the microscopic processes leading to strong turbulence in 1D superfluids, we will develop a kinetic theory of dark-soliton gases based on the Klimontovich approach \cite{schram_1991, dufty_2005}, well-established in the context of plasma physics but recently applied to atomic systems \cite{mendonca_2020a}. In integrable systems, multiple-soliton solutions, consisting of well-ordered soliton lattices, exist and are described by the inverse scattering transform method \cite{gardner_1967}. In the case of randomly distributed solitons, a statistical description in terms of distribution functions is more desirable \cite{gael_2005, bulchandani_2019}. We start by reducing DS to particle-like objects of effective negative mass, and determine the Hamiltonian and the corresponding canonical structure for a dark-soliton pair. Then, we construct a distribution function describing a collection of DS in the phase space, and obtain the corresponding transport equation. Finally, by performing ensemble averages, we are able to obtain the mean-field dynamics of a DS gas and compute its excitation spectrum. We find that a gas of dark-solitons sustains a collective excitation, which is acoustic-like (massless) in the long-wavelength limit, in agreement with the Bethe ansatz solution of the Lieb-Liniger model. Our results open a venue towards a theoretical framework able to capture the spectral properties of strong turbulence in 1D systems.    

\section{Dark-soliton Hamiltonian}

We start by considering a homogeneous, one-dimensional superfluid at zero temperature, which is governed by the Gross-Pitaevskii equation \citep{2003Pitaevskii}
\begin{equation}
i \hbar \frac{\partial \Psi (x,t)}{\partial t} = \left( - \frac{\hbar^2}{2m}\frac{\partial^2}{\partial x^2}   + g |\Psi(x,t)|^2\right) \Psi(x,t),
\label{eq_GP}
\end{equation}
with $\Psi(x,t)$ being the superfluid order parameter, associated to the following Hamiltonian density
\begin{equation}
\mathcal{H}\{\Psi\}=-\frac{\hbar^2}{2m}\left \vert \frac{\partial \Psi}{\partial x}\right \vert^2 + \frac{g}{2}\left \vert \Psi \right \vert^4.
\label{eq_ham0}
\end{equation}
Dark solitons constitute exact solutions to Eq. \eqref{eq_GP} parametrised by $s$ and $v$, standing for the centroid position and velocity, respectively, $\Psi(x,t;s,v)=e^{-i\mu t/\hbar}\psi_0[x;s,v]$, where $\mu=g n_0$ is the chemical potential, and \cite{2003Pitaevskii}
\begin{equation}
\label{eq_soliton}
\psi_{0}[x;s,v] = \sqrt{n_0} \left[ i \beta + \gamma^{-1} \tanh \left( \frac{x-s}{\gamma \xi} \right)\right].
\end{equation}
Here, $\beta=v/c$, $\gamma=(1-\beta^2)^{-1/2}$, $c=\sqrt{g n_0/m}$ is the sound speed and $\xi=\hbar/\sqrt{gmn_0}$ is the healing length. Because of the translational invariance of the solution, the soliton Hamiltonian is a function of the DS velocity only, 

\begin{equation}
H(v)=\int \left(\mathcal{H}\lbrace \Psi_0\rbrace -\mu n_0\right)dx=\frac{2}{3}mc^2 n_0\xi \left(1-\frac{v^2}{c^2}\right)^{3/2}.
\label{eq_ham1}
\end{equation}
For small velocities, $v\ll c$, $H(v)\simeq \vert M_*\vert c^2+M_*v^2/2$, where $M_*=-2mn_0\xi$ is the effective mass of the soliton \citep{1998Kivshar}. Therefore, Eq. \eqref{eq_ham1} suggests that DS may be regarded as relativistic hole-like particles, with $c$ playing here the role of the speed of light. The canonical momentum may be obtained via the relation $v=\partial_p H(v)$, which can be readily  inverted to provide
\begin{equation}
p=\int_0^v \frac{1}{u}\frac{\partial H(u)}{\partial u} du=M_*c\left(\frac{\beta}{\gamma}+\delta\right),
\label{eq_pcanon1}
\end{equation}
with $\delta=\arcsin(\beta)$ being the phase jump in the order parameter from $x-s=-\infty$ to $x-s=+\infty$. This shows that the parameter $v$ is, indeed, the kinematic DS velocity, $v=\dot s$, since $s$ and $p$ constitute a canonical pair obeying the Poisson bracket $\{s,p\}=1$. As such, Eq. \eqref{eq_pcanon1} allows us to change the functional dependence on the Hamiltonian, $H(v)\to H(p).$ \par 

With the establishment of the canonical equations for a dark soliton, we are now in position to generalize to a set of $N$ solitons. In order to avoid phase singularities within the order parameter, we consider the DS gas to be composed of an array kink$-$anti$-$kink pairs
\begin{equation}
\psi_{\rm gas}[\{s_j;v_j\}]=\prod_{j=1}^N \psi_{0}[x_j,(-1)^j v_j].
\label{eq_gasansatz}
\end{equation}
We can repeat the procedure of the single DS to formally obtain the equations of motion,
\begin{equation}
\dot s_j=\frac{\partial H_{\rm gas}}{\partial p_j}=v_j, \quad \quad \dot p_j=-\frac{\partial H_{\rm gas}}{\partial s_j},
\label{eq_ham2}
\end{equation}
with $H_{\rm gas}$ obtained from Eq. \eqref{eq_ham1} as
\begin{equation}
H_{\rm gas}(s_j,p_j)=\sum_{k\neq j}H(s_k,p_k).
\label{eq_ham3}
\end{equation}
Notice that the Hamiltonian is now a function of the soltion positions $s_j$ and their momenta $p_j$, as a consequence of the breakdown of translational invariance for the case of randomly distributed soltions. Of course, this results in $2N$ coupled equations, which are of little use. Crucially, Eq. \eqref{eq_ham2} now encodes the motion that a single soliton undergoes due to its interaction with all the others. In what follows, we implement Klimontovich procedure based on the dynamics of Eq. \eqref{eq_ham2}. 

\section{Microscopic phase-space distribution function}

In order to construct a statistical description of DS gases, we define the phase-space distribution function
of the canonical variables $s$ and $p$ as \cite{nicholson_book, stix_book}
\begin{equation}
\rho(s,p,t)=\sum_{j=1}^N\delta(s-s_j(t))\delta(p-p_j(t)),
\end{equation}
satisfying the following relation with the total number of solitons in the gas
\begin{equation}
N=\iint  \rho(s,p,t)~ds dp.
\end{equation}
Computing the time derivative, we have
\begin{equation}
\frac{\partial \rho}{\partial t}=\sum_{j=1}^N \dot s_j \frac{\partial \delta(s-s_j)}{\partial s_j}\delta(p-p_j)+\dot p_j \frac{\partial \delta(p-p_j)}{\partial p_j}\delta(s-s_j).
\end{equation}
By using the property $\partial_x\delta(x-y)=\partial_y\delta(y-x)$, we can obtain
\begin{equation}
\frac{\partial \rho}{\partial t}+\dot s \frac{\partial \rho}{\partial s}+\dot p \frac{\partial \rho}{\partial p}=0,
\end{equation}
which means that the DS phase-space distribution function may be regarded as an incompressible fluid, $\dot \rho=0$, in agreement with Liouville's theorem. The key point now is to understand that we can go from the discrete dynamics to a continuous description in the phase space by making use of $\rho(s,p,t)$, and write Eq. \eqref{eq_ham2} as $\dot p= -\partial_s V$, where $V$ is obtained from Eq. \eqref{eq_ham3} as
\begin{equation}
V(s,p,t)=\iint H_{\rm gas}(s-s',p-p')\rho(s',p',t)ds' dp'.
\label{eq_coarsepot}
\end{equation}
Notice that Eq. \eqref{eq_coarsepot} defines a pseudo-potential depending on both $s$ and $p$ (or $v$), which is a consequence of the relativistic nature of DS: their mass depends on their velocity. Together with Eq. \eqref{eq_pcanon1}, establishing a relation between $p$ and $v=\dot s$, we can recast the Klimontovich equation as
\begin{equation}
\frac{\partial \rho}{\partial t}+v \frac{\partial \rho}{\partial s}-\frac{\partial V}{\partial s} \frac{\partial \rho}{\partial p}=0. 
\label{eq_klimontovich}
\end{equation}
This approach is formally equivalent to the single-particle Liouville equation. It is very useful for numerical simulations \cite{birdsall1985plasma, pang1997introduction, surendra_1991}, but quite complicated to handle analytically. To describe the mean-field behaviour of DS gases, we introduce ensemble averages $f\equiv \langle \rho \rangle $ and $\langle V \rangle$ and the corresponding deviations as
\begin{equation}
\delta \rho=\rho -f, \quad \quad \delta V=V-\langle V \rangle,
\end{equation}
with $\langle V \rangle$ obtained from Eq. \eqref{eq_coarsepot} via the replacement $\rho \to f$. Inserting in Eq. \eqref{eq_klimontovich}, we obtain a kinetic equation for the smooth distribution function $f(s,p,t)$,
\begin{equation}
\left(\frac{\partial }{\partial t}+v \frac{\partial }{\partial s}\right)f-\frac{\partial \langle V\rangle}{\partial s} \frac{\partial f}{\partial p}=\Big\langle \frac{\partial \delta V}{\partial s} \frac{\partial \delta \rho}{\partial p}\Big\rangle. 
\label{eq_average1}
\end{equation}
The r.h.s of the latter defines the collision integral, which depends on the details of the short-range nature of DS collisions. It can be constructed, at different levels of approximations, by making use of the BBGKY hierarchy \cite{nicholson_book}, thus yielding different kinetic equations. In what follows, we consider dilute soliton gases, $N_0\xi\ll 1$, where $N_0=1/\langle s\rangle$ is the gas density determined by the averaged soliton separation $\langle s\rangle $. In that regime, we neglect the correlations between the multi-particle distributions and set the collision integral in Eq. \eqref{eq_average1} to zero. As such, we obtain the collisionless kinetic equation for the single-particle distribution function 
\begin{equation}
\left(\frac{\partial }{\partial t}+v \frac{\partial }{\partial s}\right)f-\frac{\partial \langle V\rangle}{\partial s} \frac{\partial f}{\partial p}=0. 
\label{eq_vlasov}
\end{equation}
\begin{figure}[t!]
\includegraphics[width=1.\columnwidth]{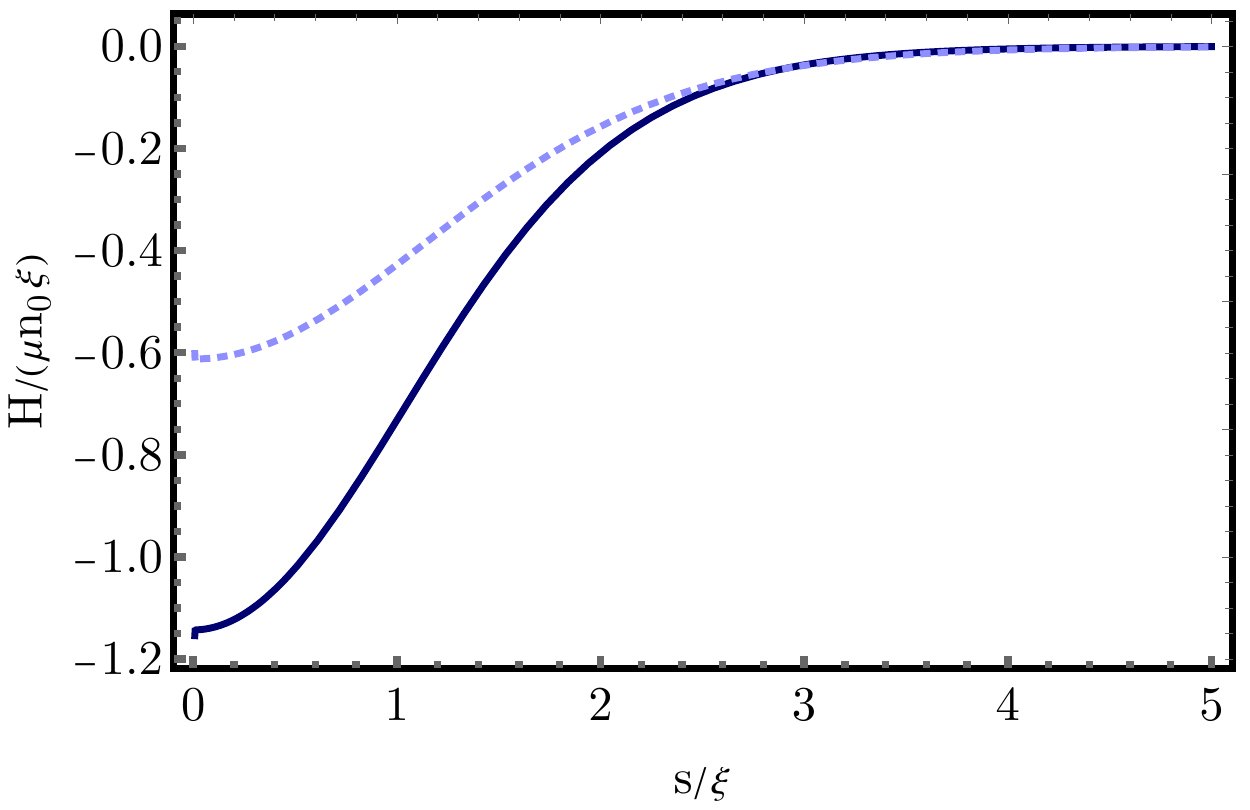}
\caption{The effective pairwise pseudo-potential as a function of the soliton separation for $v=0$ (solid line) and $v=0.5 c$ (dashed line).}
\label{fig1}
\end{figure}
The latter is formally equivalent to the Vlasov equation widely used to describe fully ionised plasmas \cite{stix_book, vlasov_1961}. The Vlasov equation has also been employed to describe photon-quasiparticles, where long-range interactions are absent \cite{mendonca_2014a}. As we are about to see, the important difference stems from the fact that the mean-field soliton-soliton interaction is short-ranged, contrary to the case of electrons and ions in plasmas interacting through the Coulomb potencial. \par
In order to model the interaction potential in Eq. \eqref{eq_vlasov}, we consider that DS interact via an averaged two-body potential. This is justified since dark solitons are localised objects of size $\sim \xi$, making their interaction to be finite-ranged \cite{frantzeskakis_2010}. As such, the two-body pseudo-potential is obtained from Eqs. \eqref{eq_gasansatz} and \eqref{eq_ham3} by setting $N=2$, and therefore defining $H_{\rm gas}\simeq H_{\rm pair}\equiv H(s,v)$, where
\begin{equation}
H(s,v)\simeq H_{\rm kin}(s,v) +H_{\rm int}(s,v)-\frac{8}{3}\vert M_*\vert c^2\left(1-\frac{v^2}{c^2}\right)^{-3/2}.
\label{eq_pairwise}
\end{equation}
Here,
\begin{equation}
\begin{array}{ccl}
H_{\rm kin}(s,v) &=&-\frac{1}{3}\frac{gn_0^2\xi}{\gamma^{3/2}} \left[2 \beta ^2-\cosh \left( \zeta \right)-1 \right] \\\\ 
&\times&\text{csch}\left( \zeta \right)
\left[\cosh \left(2\zeta\right)-  6\zeta \coth \left( \zeta\right)+5\right]
\end{array}
\end{equation}
is the contribution from the kinetic term in Eq. \eqref{eq_ham0} and
\begin{equation}
\begin{array}{ccl}
H_{\rm int}(s,v) &=& \frac{1}{48}\frac{gn_0^2\xi}{\gamma^{3/2}} \text{csch}^7(\zeta) \left[\left(16 \beta ^2-35\right) \sinh \left(5\zeta\right)\right. \\\\
&+&\left.\sinh \left(7\zeta\right)+12 \zeta \cosh \left(5\zeta\right) \right.\\\\
&+& \left.\left(-432 \beta ^4+448 \beta ^2-171\right) \sinh \left(\zeta\right)\right. \\\\
&+& \left. \left(-176 \beta ^4+464 \beta ^2-207\right) \sinh \left(3\zeta \right) \right. \\\\
&+& \left. 24 \left(36 \beta ^4-64 \beta ^2+29\right) \zeta \cosh \left(\zeta\right)\right. \\\\
&+& \left. 12 \left(8 \beta ^4-32 \beta ^2+21\right) \zeta \cosh \left(3\zeta\right)\right],  
\end{array}
\end{equation}
results from the interaction (non-linear) term, where $\zeta=s/(\gamma \xi)$, $s$ is the soliton separation and $v=(v_1+v_2)/2 $ is the average velocity of the soliton pair. The third term in Eq. \eqref{eq_pairwise} corresponds to twice the energy of a single soliton, $s\to \infty$, and does not contribute to the force term in the Vlasov equation.
As it can be seen in Fig. \eqref{fig1}, the pairwise DS potential is attractive. However, since the DS mass is negative (and hence the reduced mass of the DS pair), the resulting interaction is repulsive. Moreover, we can observe that the potential has a range of order $\sim \xi$, and becomes weaker for more relativistic (less massive, in modulus) solitons.

\section{Sound modes of the dark-soliton gas}

In order to illustrate some of the features of the transport equation, we start by considering small amplitude perturbations around a certain equilibrium configuration, 
\begin{equation}
f\simeq f_0+\tilde f_1, \quad {\rm with} \quad f_1\ll f_0.
\end{equation}
Inserting this into Eq. \eqref{eq_vlasov}, we obtain the linearised Vlasov equation
\begin{equation}
\left(\frac{\partial}{\partial t}+v\frac{\partial}{\partial s}\right)f_1-\frac{\partial \langle V_1\rangle}{\partial s}\frac{\partial f_0}{\partial p}=0,
\end{equation}
where $\langle V_1(s,p)\rangle =\int H(s-s',p-p')f_1(s',p')ds'dp'$ with $H$ being given by Eq. \eqref{eq_pairwise}. Upon Fourier transforming the latter (i.e. by making $f_1(s,p,t)=\sum_{k,\omega}\tilde f_1(k,p,\omega) e^{i(ks-\omega t)}$, we formally obtain the kinetic dispersion relation of the DS gas as
\begin{equation}
1=k\int \frac{\tilde H(k,v)}{(\omega- kv)}\frac{\partial f_0}{\partial v}\frac{\partial v}{\partial p}~dv,
\end{equation}
where $\tilde  H(k,v)$ is the spatial Fourier transform of Eq. \eqref{eq_pairwise}. The Jacobian $\partial v/\partial p$ allows us to eliminate $p$ and corresponds to a generalised mass term  that can be determined with the help of Eq. \eqref{eq_pcanon1}. The dispersion relation can be numerically solved for generic equilibrium configurations, accounting for i) the relativistic nature of dark solitons, ii) the velocity-dependence of their pairwise interaction, and iii) the negativity of their mass (``hole-like" nature). A particularly interesting and analytically tractable case is that of a non-relativistic DS gas, distributed such that $v\ll c$. In that case, we can set $H(k,v)\simeq \tilde H(k,0)$ and $\partial p/\partial v\simeq-1/\vert M_*\vert$, which  yields
\begin{equation}
1\simeq- \frac{k \tilde H(k,0)}{\vert M_*\vert }\int \frac{1}{(\omega- kv)}\frac{\partial f_0}{\partial v}~dv,
\label{eq_dispersion}
\end{equation}
where 
\begin{equation}
\tilde H(k,0)\simeq -\frac{1}{2}\vert M_*\vert c_s^2\xi \left( \frac{28}{9}+(7\pi^2-15)k^2\xi^2\right)+\mathcal{O}(k^4).
\label{eq_fourier}
\end{equation}
As we can immediately see, the negative signs in Eqs. \eqref{eq_dispersion} and \eqref{eq_fourier} cancel, thus confirming that the nature of the DS interaction is, indeed, repulsive as earlier stated. \par
We consider fluctuations on top of homogeneous soliton gases, $f_0(s,v)=N_0 g_0(v)$. An interesting situation corresponds to that of a cold dark-soliton gas, for which solitons are prepared at rest, distributed in velocity as $g_0(v)=\delta(v)$. Although this distribution is correct for classical particles, it may not be generally accurate to describe solitons. The reason stems from the fact that solitons are fermions, as one can immediately see from Eq. \eqref{eq_gasansatz} for a two-DS wavefunction \cite{tercas_2013},
\begin{equation}
\psi_{2}(x_2,x_1;0,0)=-\psi_{2}(x_1,x_2;0,0),
\end{equation}
in agreement with Lieb-Liniger theory \cite{sato_2012, karpiuk_2015}. The fermionisation of the DS gas must therefore be included in the equilibrium at a semi-classical level. For that task, we make use of the distribution function
\begin{equation}
g_0(v)=\frac{1}{2v_F}\Theta(v_F-\vert v\vert), 
\end{equation}
where $v_F=\pi\hbar N_0/\vert M_*\vert$ is the 1D Fermi velocity of the DS gas. As a matter of fact, the later reduces to the Dirac-delta distributed gas in the limit $v_F\to 0$, i.e. for sufficiently diluted DS gases. The dispersion relation in \eqref{eq_dispersion} thus provides 
\begin{equation}
\omega^2=v_F^2k^2+\frac{N_0 k^2\tilde H(k,0) }{M_*}
\label{eq_modes1}
\end{equation}
In the long-wavelength limit $k\xi\to 0$, this corresponds to a {\it solitonic first sound}, $\omega \simeq v_1 k$, where 
\begin{equation}
v_1=c \sqrt{\frac{\pi^2}{4}\Gamma^2+\frac{14}{9\pi}\Gamma}
\end{equation}
is the first sound speed and $\Gamma=N_0\xi$ is the DS gas dilution parameter. In the validity range of Eq. \eqref{eq_vlasov}, $\Gamma\ll 1$, $v_1/c\simeq 0.703 \sqrt{\Gamma}$, meaning that the effects of the Fermi pressure are not relevant for very dilute gases. Indeed, the effect of the fermionisation becomes relevant for densities of the order $\Gamma\gtrsim (56/(9\pi^3))\simeq 0.203$.  \par
\begin{figure}[t!]
\includegraphics[width=1.\columnwidth]{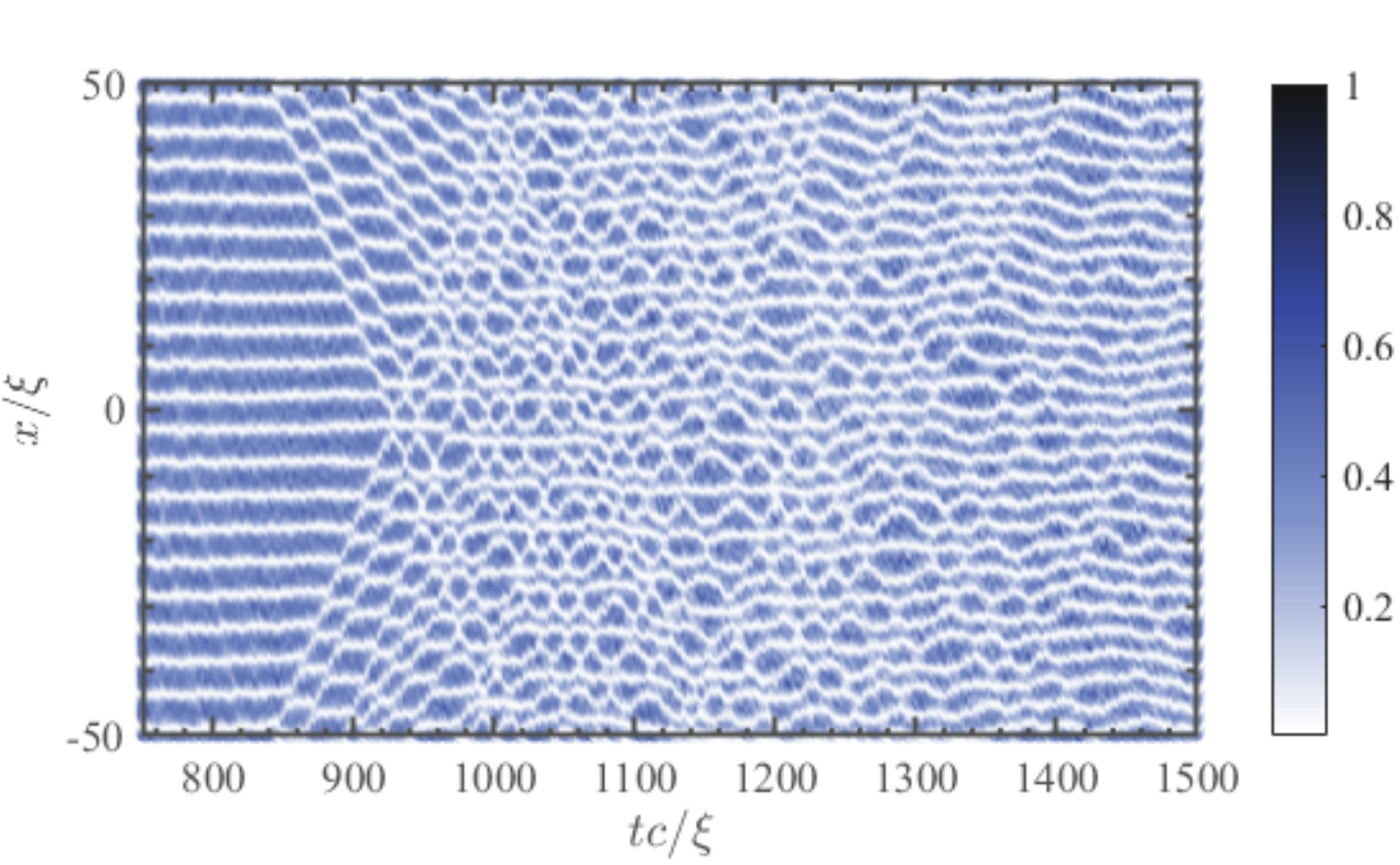}
\caption{Numerical simulation of a cold dark-soliton gas, $g_0(v)=\delta(v)$, depicted for a concentration parameter $\Gamma=0.2$. One observes the formation of an acoustic mode, which preludes the solitonic turbulence.}
\label{fig_gas_real}
\end{figure}
In order to validate the kinetic approach developed here, we have integrated Eq. \eqref{eq_GP} with the initial condition of Eq. \eqref{eq_gasansatz}, making use of a time-splitting finite difference method \cite{antoine_2013}. As we can observe in Fig. \ref{fig_gas_real}, the DS gas seems to sustain a collective mode which is acoustic (non-dispersive) in the long wavelength limit. In order to characterize this mode quantitatively and to compare with the theory, we make use of the dynamical structure factor 
\begin{figure}[t!]
\includegraphics[width=1.\columnwidth]{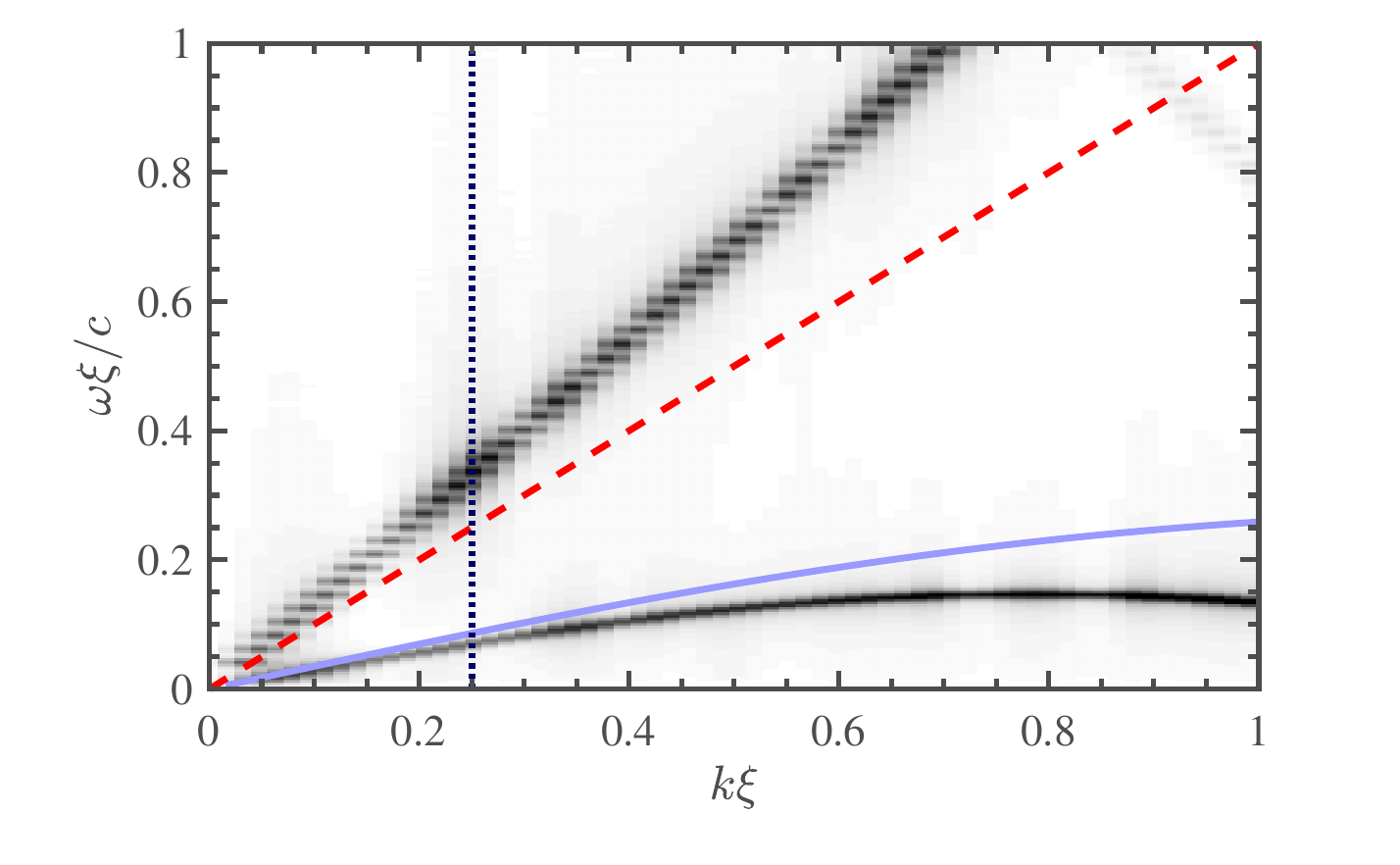}
\includegraphics[width=1.\columnwidth]{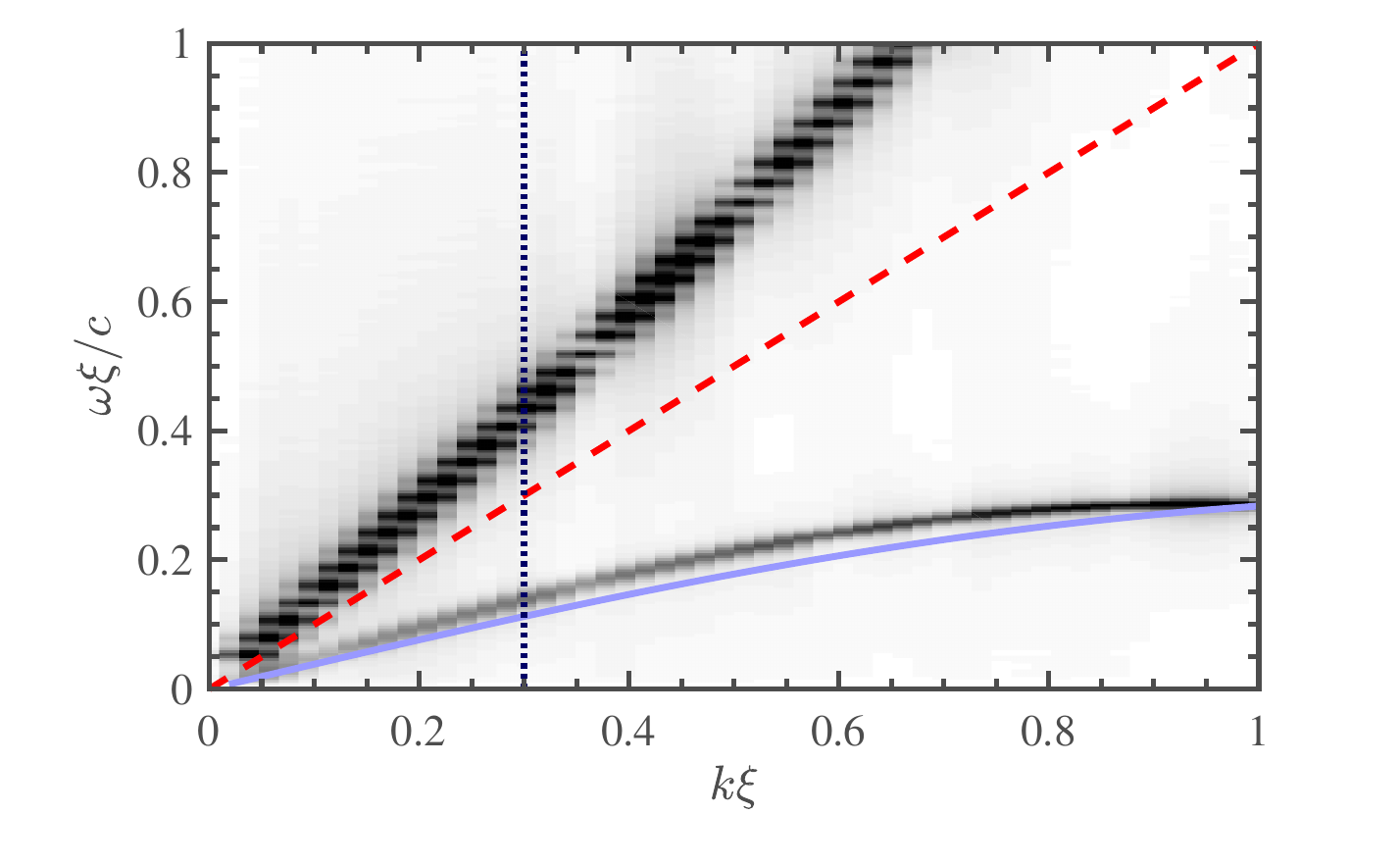}
\caption{Dynamic structure factor $S(\omega,k)$ as obtained from the numerical integration of Eq. \eqref{eq_GP}, for $\Gamma=0.25$ (top) and $\Gamma=0.3$ (bottom). In both panels, the solid line corresponds to the first-sound mode in Eq. \eqref{eq_modes1}, plotted here with no adjustable parameters, while the top curve indicates an hybridization mode between Bogoliubov excitations (dashed lines) and DS first sound with dispersion $\omega\simeq v_2k$ (not shown, see text). The vertical dotted lines depicts the border of the expected validity of the Vlasov equation \eqref{eq_vlasov}, $k= N_0$.  }
\label{fig_dynamicfactor}
\end{figure}
\begin{equation}
S(\omega, k)=\iint e^{-i\omega t+ik x}\vert \Psi (x,t)\vert^2 dx dt,
\end{equation}
as depicted in Fig. \ref{fig_dynamicfactor}. As we can see, there is a good agreement between the first sound mode in Eq. \eqref{eq_modes1} and the numerics is obtained in the region $k \lesssim N_0$, i.e. for wavelengths lying above the inter-particle separation $1/N_0$. Above this value, the coarse-graining assumption of the phase space breaks down. Moreover, for shorter wavelengths in the range $k\sim \xi $, the description of DS as hole-like particles fails and the internal structure of the solitons becomes important, what hinders the validity of the kinetic equation. Finally, the effects of collision integral in Eq. \eqref{eq_average1} are expected to play a prominent role for sufficiently dense DS gases $N_0\sim \xi $, as well as for sufficiently large velocities, $v\sim c$. In both situations, the short-range binary collisions between solitons needs to be taken into account. \par
An additional feature can be observed in Fig. \eqref{fig_dynamicfactor}: the emergence of an energetic mode $\omega\simeq v_2 k$, with $v_2>c$. This mode does not correspond to the low-lying (Bogoliubov) excitations on top of the superfluid - a fact that we verified numerically - and is certainly not that of a soliton gas (for which the slope is $v_1\ll c$, as discussed above). At this stage, we understand this mode to be a consequence of the hybridization between the Bogoliubov (fast) and the first-sound (slow) modes, as a result of the interactions between solitons and phonons. Although a microscopic theory remains to be developed, we can anticipate that our kinetic theory may be coupled to a Wigner-Moyal description of phonons \cite{mendonca_book}. The resulting problem will consist on the dynamics under the effect of a ponderomotive force due to the modulations of the superfluid density by the phonons \cite{cary_1981, SHUKLA_19861}.

\section{Conclusions}

We have established the foundations of a kinetic theory of dark soliton gases in one-dimensional superfluids based on the Klimontovich approach. By considering that dark solitons behave as particles of negative mass, and assuming that they interact via an ensemble averaged pairwise potential, we defined a kinetic equation governing the phase-space distribution function of an array of dark solitons, i.e. a dark soliton gas. Our approach allows us to describe solitons statistically in analytical grounds, a feature that is virtually impossible with studies based on the Gross-Pitaevskii equation. One important feature of the soliton gas is that it supports a first sound mode, which is much less energetic than the phonon modes on top of the condensate. This as consequence of solitons behaving as weakly interacting particles of negative mass. We believe to be setting the stage towards a more comprehensive, microscopic description of solitonic turbulence, an aspect of central importance when dealing with strong turbulence regimes in one-dimensional superfluids. In a near future, with the help of a hydrodynamic model that can be directly obtained from our kinetic equation, we believe to be able to describe solitonic turbulence as a Kolmogorov theory of weak turbulence of dark soliton gases. If successful, such a description will unveil important mechanisms underlying the spectral properties of strong quantum turbulence in low-dimensional superfluids.

\acknowledgments

The authors acknowledge Funda\c{c}\~{a}o para a Ci\^{e}ncia e Tecnologia (FCT-Portugal) through the Contract No. CEECIND/00401/2018 and the financial support from the Quantum Flagship Grant PhoQuS (Grant No. 820392) of the European Union.

\bibliographystyle{apsrev4-1}
\bibliography{REFERENCES}
\bibliographystyle{apsrev4-1}

\end{document}